\documentclass[a4paper,11pt]{article}
\pdfoutput=1 % if your are submitting a pdflatex (i.e. if you have
\usepackage{jcappub} % for details on the use of the package, please
\usepackage[T1]{fontenc} % if needed
\usepackage[utf8]{inputenc}
\usepackage{lmodern}

\usepackage[english]{babel}
\usepackage{booktabs}
\usepackage{amsmath}
\usepackage{wasysym}
\usepackage{graphicx}
\usepackage{esint}

\def\rd{{\rm d}}

\definecolor{vale}{rgb}{0,0.5, 1.}

\title{Clustering dark energy and halo abundances}

\author[a]{Ronaldo C. Batista}
\author[b]{and Valerio Marra}

\affiliation[a]{Escola de Ci\^encias e Tecnologia, Universidade Federal do Rio
Grande do Norte\\
Campus Universitário Lagoa Nova, Natal, RN, Brazil, CEP 59078-970\\
and\\
International Institute of Physics, Universidade Federal do Rio
Grande do Norte\\
Campus Universitário Lagoa Nova, Natal, RN, Brazil, CEP 59078-970
}
\affiliation[b]{Departamento de Física, Universidade Federal do Esp\'{\i}rito 
Santo\\
Av.~F.~Ferrari, 514, 29075-910, Vit\'oria, ES, Brazil}

\emailAdd{rbatista@ect.ufrn.br}
\emailAdd{valerio.marra@me.com}

\abstract{
Within the standard paradigm, dark energy is taken as a homogeneous fluid that 
drives the accelerated expansion of the universe and does not contribute to 
the mass of collapsed objects such as galaxies and galaxy clusters. The 
abundance of galaxy clusters -- measured through a variety of channels -- has 
been extensively used to constrain the normalization of the power spectrum: it 
is an important probe as it allows us to test if the standard $\Lambda$CDM model 
can indeed accurately describe the evolution of structures across billions of 
years. It is then quite significant that the Planck satellite has detected, via 
the Sunyaev-Zel'dovich effect, less clusters than expected according to the 
primary CMB anisotropies. One of the simplest generalizations that 
could reconcile these observations is to consider models in 
which dark energy is allowed to cluster, i.e., allowing its sound speed to 
vary. In this case, however, the standard methods to compute the abundance of 
galaxy clusters need to be adapted to account for the 
contributions of dark energy. In particular, we examine the case of 
clustering dark energy -- a dark energy fluid with negligible sound speed -- 
with a redshift-dependent equation of state. We carefully study how the halo 
mass function is modified in this scenario, highlighting corrections that have 
not been considered before in the literature. We address modifications in the 
growth function, collapse threshold, virialization densities and also changes in 
the comoving scale of collapse and mass function normalization.
Our results show that clustering dark energy can impact halo abundances at the 
level of 10\%--30\%, depending on the halo mass, and that cluster counts are 
modified by about 30\% at a redshift of unity.

}

\begin{document}
\maketitle
\flushbottom

\section{Introduction}

The abundance of galaxy clusters is a powerful tool to
study the late time evolution of the universe. These objects are formed
at low redshifts and strongly depend on the properties of dark matter,
baryons and the physics of cosmic acceleration.
Analyses of cluster counts depend crucially on the knowledge of the halo mass
function.
Thanks to the approximate universality of the mass function, the latter can be 
calibrated once and for all using large-scale high-resolution $N$-body 
simulations~\cite{Tinker2008b}.
However, these computationally very expensive simulations address only the 
simple standard $\Lambda$CDM model or, at most, a smooth dark energy (DE) with 
constant equation of state parameter (EoS), $w=\bar{p}_{de}/\bar{\rho}_{de}$.
This limits the possible analyses of cluster counts to these very standard 
models.

Cluster counts have been extensively used to constraint the matter density 
parameter $\Omega_m$ and the normalization $\sigma_8$ of the power spectrum, 
see, e.g., \cite{Rozo:2007yt,Mantz2008,Vikhlinin:2008ym,Ade:2015fva}.
The abundance of clusters is an important probe in cosmology as it allows us to 
test the evolution of structures across billions of years.
It was then unexpected that the Planck satellite detected via the 
Sunyaev-Zel'dovich effect less clusters than predicted according to the primary 
CMB anisotropies~\cite{Ade:2015fva}.
This tension could be due to a poor knowledge of the scaling relation 
calibration or to a non-standard energy component in the universe. The latter 
possibility motivates us to extend state-of-the-art mass functions to the case 
of a dark energy that can cluster and so participate in the formation of 
collapsed objects such as galaxies and galaxy clusters.
The new mass function can then be used to analyze the Planck clusters and see 
if a clustering dark energy can help in alleviating the tension from 
Planck~\cite{RoVaX}.

Within General Relativity, the simplest extension with respect to the 
cosmological constant $\Lambda$ is a dark energy with a time varying equation
of state. In this case, DE fluctuations are inevitably present, giving rise to 
at least a new degree of freedom: the sound speed (in the rest frame of
the fluid) $c_{s}=\sqrt{\delta p_{de}/\delta\rho_{de}}$. In quintessence
models, DE is represented by a canonical scalar field with a potential,
which determines the evolution of the EoS in the range $-1<w\left(t\right)<1$.
However, quintessence models always have $c_{s}=1$. This fact implies
that quintessence perturbations are very small compared to matter
perturbations on small scales (see though~\cite{Amin2012a}).
This gives support to the common practice of neglecting DE perturbations
in structure formation studies such as $N$-body simulations. 

However, there exists many DE models which can be described by a perfect
fluid with varying sound speed. Models that show this behavior include
the tachyon field \cite{Padmanabhan:2002cp,Bagla:2002yn} and, in
general, the whole class of minimally coupled $K$-essence scalar 
fields~\cite{Garriga:1999vw,Armendariz-Picon2001}, in which
the sound speed can be freely chosen. Models with negligible sound speed
can also be constructed with two scalar fields \cite{Lim2010} and
effective field theory \cite{Creminelli2009}.

When DE has negligible sound speed, its perturbations have no pressure
support and can grow at the same pace as matter perturbations. Therefore, they 
can evolve nonlinearly and impact the formation
of galaxy clusters.
As cosmological simulations that include
DE fluctuations have not yet been conducted, one has to resort to 
semi-analytical approaches
such the spherical collapse model (SCM) \cite{Gunn:1972sv} together
with Press-Schechter (PS) or Sheth-Tormen (ST) mass functions 
\cite{Press:1973iz,Sheth:1999su} in order to explore the impact of clustering 
DE in the abundances of halos.
In fact, the SCM was generalized to study clustering
DE and it was shown that DE fluctuations can indeed become nonlinear,
impact matter growth and the critical density threshold for collapse,
$\delta_{c}$, and contribute to the mass of the clusters 
\cite{Mota:2004pa,Manera:2005ct,Abramo2007,Abramo2009,Creminelli2010,Basse2011,
Batista:2013oca,Pace2014a,Heneka:2017ffk}. The impact of clustering DE was also 
studied in higher order perturbation theory
\cite{Sefusatti2011,Fasiello2016,Fasiello2016a}.

In this work we carefully examine how the halo mass function is modified in 
the presence of clustering DE -- a dark energy fluid with negligible sound 
speed -- with a redshift-dependent equation of state, highlighting corrections 
that have not been considered in the literature before. 
We reanalyze some modifications that were already considered in the literature, 
such as the growth function, collapse threshold and virialization densities. 
Moreover, we also consider two new modifications: the 
comoving scale of collapse and mass function normalization.
Section~\ref{scm} is devoted to the study of the SCM in the case of clustering 
DE, while Section~\ref{shmf} discusses how the halo mass function is modified 
in order to take into account the contribution of DE fluctuations to the halo 
virialization. In Section~\ref{results} we show that clustering dark energy can 
impact halo abundances at the level of 10\%--30\%, depending on the halo mass, 
and that cluster counts are modified by about 30\% at a redshift of unity. We 
draw our conclusions in Section~\ref{conclusions}.

Throughout this work we adopt the following fiducial values of the cosmological 
parameters: $h=0.7$, $\Omega_m=0.3$, $\Omega_k=0$, $\Omega_b \, h^2 = 0.0223$, 
$\sigma_8=0.8, n_s=0.966$.

\section{Spherical collapse in the fluid and radius approach} \label{scm}

The SC model can be generalized to treat fluids other than pressureless
matter using the Pseudo-Newtonian Cosmology 
approach~\cite{Abramo2007,Abramo2009,Creminelli2010}.
This framework is particularly useful to study clustering DE ($c_{s}=0$),
in which case the equations are:
\begin{equation}
\dot{\delta}_{m}+3\frac{\theta}{a}\left(1+\delta_{m}\right)=0\,,
\label{eq:delta_m_fluid}
\end{equation}
\begin{equation}
\dot{\delta}_{de}-3wH\delta_{de}+3\frac{\theta}{a}\left(1+w+\delta_{de}
\right)=0\,,
\label{eq:delta_e_fluid}
\end{equation}
\begin{equation}
\dot{\theta}+H\theta+\frac{\theta^{2}}{3a}=-4\pi 
Ga\left(\bar{\rho}_{m}\delta_{m}+\bar{\rho}_{de} 
\delta_{de}\right)\,,\label{eq:theta_fluid}
\end{equation}
where $\theta$ is the divergence of the peculiar velocity, 
$\delta_{m}=\delta\rho_{m}/\bar{\rho}_{m}$
and $\delta_{de}=\delta\rho_{de}/\bar{\rho}_{de}$. In this approach,
the redshift of collapse, $z_{c}$, is determined when 
$\delta_{m}\rightarrow\infty$,
which is numerically implemented as certain threshold that reproduces
the EdS results. As we will show, although this method is widely
used to study the nonlinear evolution, the determination of the critical
threshold might not be the best approach when DE fluctuations are present. 

In order to understand the issue, let us now consider the radial approach.
The original SC model, which determines the evolution of a spherical
shell of physical radius $R$, only considers pressureless matter
and can be solved analytically. When DE with negligible $c_{s}$ is
present, the dynamical equation for $R$ can be written as
\begin{equation}
\frac{\ddot{R}}{R}=-\frac{4\pi 
G}{3}\left[\bar{\rho}_{m}\left(1+\delta_{m}\right)+\bar{\rho}_{de}
\left(1+3w+\delta_{de}\right)\right]\,.\label{eq:radius}
\end{equation}
We also need equations for the contrasts of matter and DE. For matter,
we assume that the total mass inside the shell with radius $R$ is
conserved, then we have 
\begin{equation}
\dot{\delta}_{m}+3\left(\frac{\dot{R}}{R}-\frac{\dot{a}}{a}
\right)\left(1+\delta_{m}\right)=0.\label{eq:matter_constrast}
\end{equation}
Comparing with Eq.~(\ref{eq:delta_m_fluid}), we identify 
\begin{equation}
\frac{\theta}{a}=3\left(\frac{\dot{R}}{R}-\frac{\dot{a}}{a}\right).
\label{eq:theta_radius}
\end{equation}
Note that deriving Eq.~(\ref{eq:theta_radius}) with respect to time and using 
Eq.~(\ref{eq:radius}),
we recover Eq.~(\ref{eq:theta_fluid}). The radial approach itself can not 
determine the evolution of DE fluctuations -- note that the evolution of matter 
fluctuations is given by the assumption that the total matter within $R$ is 
conserved, which is not valid for DE. It is possible to derive an equation for 
DE fluctuations using the relation between
$\theta$ and $R$, and Eq.  (\ref{eq:delta_e_fluid}), which yields
\begin{equation}
\dot{\delta}_{de}-3wH\delta_{de}+3\left(\frac{\dot{R}}{R}-\frac{\dot{a}}{a}
\right)\left(1+w+\delta_{de}\right)=0.
\label{eq:DE_contrast}
\end{equation}
Hence, the system of Eqs.~(\ref{eq:radius}), (\ref{eq:matter_constrast})
and (\ref{eq:DE_contrast}) determines the evolution of the shell
radius, matter and DE fluctuations. The dynamical evolution in the two 
approaches is identical, however, as we will show, the commonly used collapse 
criteria in the two approaches are not equivalent.  

In the radius approach, the redshift of collapse, $z_{c}$, is given
when $R\rightarrow0$. In the EdS model, the RHS of Eq.~(\ref{eq:radius}) only 
has matter quantities and the critical density of collapse takes the standard 
value $\delta_{c}=\delta_{m}^{{\rm lin}}\left(z_{c}\right)\simeq1.686$.
In the presence of DE, however, Eq.~(\ref{eq:radius}) shows that DE 
fluctuations also contribute to the collapse, hence the 
density contrast that gives the threshold density contrast has to be modified 
to take this new contribution into account. Therefore, instead of determining
the collapse threshold only with the matter contribution, 
the natural quantity to use is the weighted total fluctuation: 
\begin{equation}
\delta_{{\rm tot}}\left(z\right)=\delta_{m}\left(z\right)+\frac{\Omega_{de}
\left(z\right)}{
\Omega_{m}\left(z\right)}\delta_{de}\left(z\right).
\label{eq:delta_tot}
\end{equation}
At high-z, when DE is subdominant in the background, the contribution of DE 
fluctuations to $\delta_{{\rm tot}}$ is very small, regardless of the 
magnitude of $\delta_{de}$. Conversely, at low-z, DE dominates the background 
and even relatively small DE fluctuations impact $\delta_{{\rm tot}}$. 
So, the definition of $\delta_{{\rm tot}}$ reflects the fact the gravitational 
potential, which drives the evolution of $R$, is sourced by densities 
fluctuations, $\bar{\rho}\delta$. The same quantify was used to study higher 
order perturbation theory in clustering quintessence models \cite{Sefusatti2011}.

The critical contrast is then 
given by the linearly evolved total fluctuation: 
\begin{equation}
\delta_{c}\left(z_{c}\right)=\delta_{{\rm tot}}^{{\rm lin}}\left(z_{c}\right) \,,
\end{equation}
where $z_{c}$ is the redshift at which $\delta_{{\rm 
tot}}$ is above a certain numerical threshold.
In the same fashion, the normalized growth function is given by
\begin{equation}
G_{{\rm tot}}\left(z\right)=\frac{\delta^{{\rm lin}}_{{\rm 
tot}}\left(z\right)}{\delta^{{\rm lin}}_{{\rm 
tot}}\left(0\right)}.
\label{eq:growth_N}
\end{equation}

Another interesting point about the relation between these two approaches
is the error in $\delta_{c}$ when using the fluid approach, as reported
in~\cite{Herrera:2017epn} (see also \cite{Pace:2017qxv}). The authors show that 
the numerical collapse threshold is not the same for different redshifts, which 
in turn generates a spurious increase
of $\delta_{c}$ with $z$. This can be corrected by calibrating the numerical
threshold as a function of the redshift in order to reproduce some known
$\delta_{c}$ and demanding that it approaches the EdS value at high-$z$.
Once this calibration is done, we verified that both approaches give
essentially the same results for the models under consideration in
this work. However, the radial approach is more robust because it
does not demand any redshift-dependent calibration and for this reason it will 
be used here.

\subsection{Background evolution and growth function} \label{bck_mods}

In order to show the impact of DE fluctuations on the halo collapse, we adopt a 
DE with a linear parametrization of the equation of state,
$w=w_{0}+\left(1-a\right)w_{a}$.
In this work we do not consider models with $w<-1$ because in such
case, during the nonlinear regime, the dynamical evolution can generate
$\delta_{de}<-1$, which implies a non-physical negative energy density. 
In fact, we observed this happens around virialization densities for $1+w \simeq 
- 0.1$. Hence we prefer to focus on the less problematic region of 
parameter space of non-phantom equations of state in order to clearly show the 
modifications in the mass function.

The latter happens because in Eq.~(\ref{eq:delta_e_fluid}) the coupling
between the density contrast and gravitational force is proportional
to $\left(1+w+\delta_{de}\right)$. Hence, already at the linear regime,
matter overdensities will create DE underdensities if $w<-1$. Indeed, in the 
matter dominated era and for constant $w$ one 
has~\cite{Abramo2009,Batista:2013oca}:
\begin{equation}
\delta_{de}=\frac{1+w}{1-3w}\delta_{m}\,. \label{eq:de_mat_relation}
\end{equation}
In the nonlinear regime, $\delta_{de}$
continues to decrease as $\delta_{m}$ grows, but the coupling does
not vanish as $\delta_{de}\rightarrow-1$ negative energy densities
can appear. To better understand this behavior, bear in mind that voids of 
matter always have $\delta_{m}>-1$ because the coupling between its density 
contrast and gravitational force is $\left(1+\delta_{m}\right)$, so as 
$\delta_{m}\rightarrow-1$ the coupling vanishes and the decrease in 
$\delta_{m}$ is halted. In the phantom case, $\left(1+w+\delta_{de}\right)$ is 
always negative inside matter halos, therefore $\delta_{de}$ continues to 
decrease as $\delta_{m}$ grows. This pathological behavior might be an indication 
of a flaw in phantom models with negligible sound speed. This issue is not 
present for $w>-1$ because in this case matter halos induce positive 
$\delta_{de}$ and, even in the
case of an extreme matter void with $\delta_{m}\simeq-1$, $\delta_{de}$
is smaller in magnitude than $\delta_{m}$ for $w\apprge-1$.

\begin{table}
\begin{center}
\begin{tabular}{ccc}
 & $w_{0}$ & $w_{a}$\tabularnewline
\midrule 
S1 & -0.9 & 0.2\tabularnewline
\midrule 
S2 & -0.9 & 0.1\tabularnewline
\midrule 
S3 & -0.9 & 0\tabularnewline
\midrule 
S4 & -0.9 & -0.1\tabularnewline
\bottomrule
\end{tabular}
\caption{The four sets of parameters $w_{0}$ and $w_{a}$ 
used in this work. \label{tab:params}}
\end{center}
\end{table}

We choose the four sets of parameters shown in Table \ref{tab:params}.
As initial conditions, we assume that only the growing mode of matter
is present and $\delta_{de}$ is given by Eq.~\eqref{eq:de_mat_relation}.

In Figure~\ref{fig:growth} we show, for each 
set of parameters, the ratios of the total growth function, 
$G_{{\rm tot},c_s=0}$ (left panel) and the usual matter only growth function, 
$G_{m,c_s=0}$, (right panel) to the matter growth function for the 
case of negligible DE perturbations, $G_{m,c_s=1}$, as a function of redshift. 
As we can see, the impact of DE perturbations on $G_{{\rm tot}}$ is much 
larger. Given this fact, we already expect that our proposed growth function 
will strongly impact the abundance of galaxy clusters. Since all functions are 
normalized at $z=0$, we can see that clustering DE enhances the growth of 
perturbations because they grow from smaller values in the past to reach the 
same value as the homogeneous case today. 

Also note that the greatest impact of DE perturbations appear for S1. This 
happens because S1 has the largest value of $w$ at high-$z$,
hence both $\delta_{de}$ and 
$\Omega_{de}\left(z\right)/\Omega_{m}\left(z\right)$
are enhanced. This kind of behavior is responsible for large deviations
of the growth factor with respect to the so called GR value \cite{Batista2014a}.

\begin{center}
\begin{figure}
\centering{}
\includegraphics[width=.48 \columnwidth]{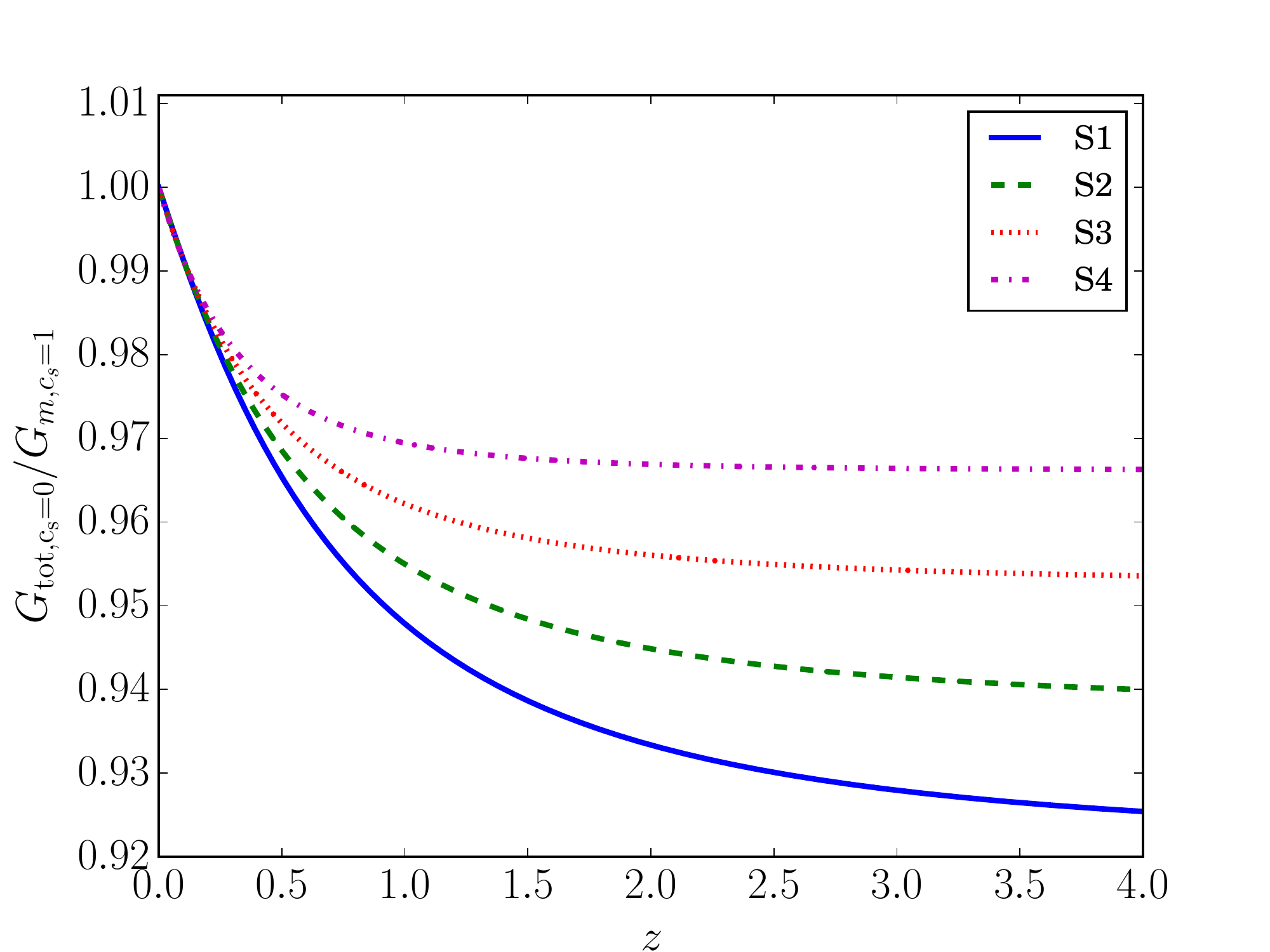}
\hfill
\includegraphics[width=.48 \columnwidth]{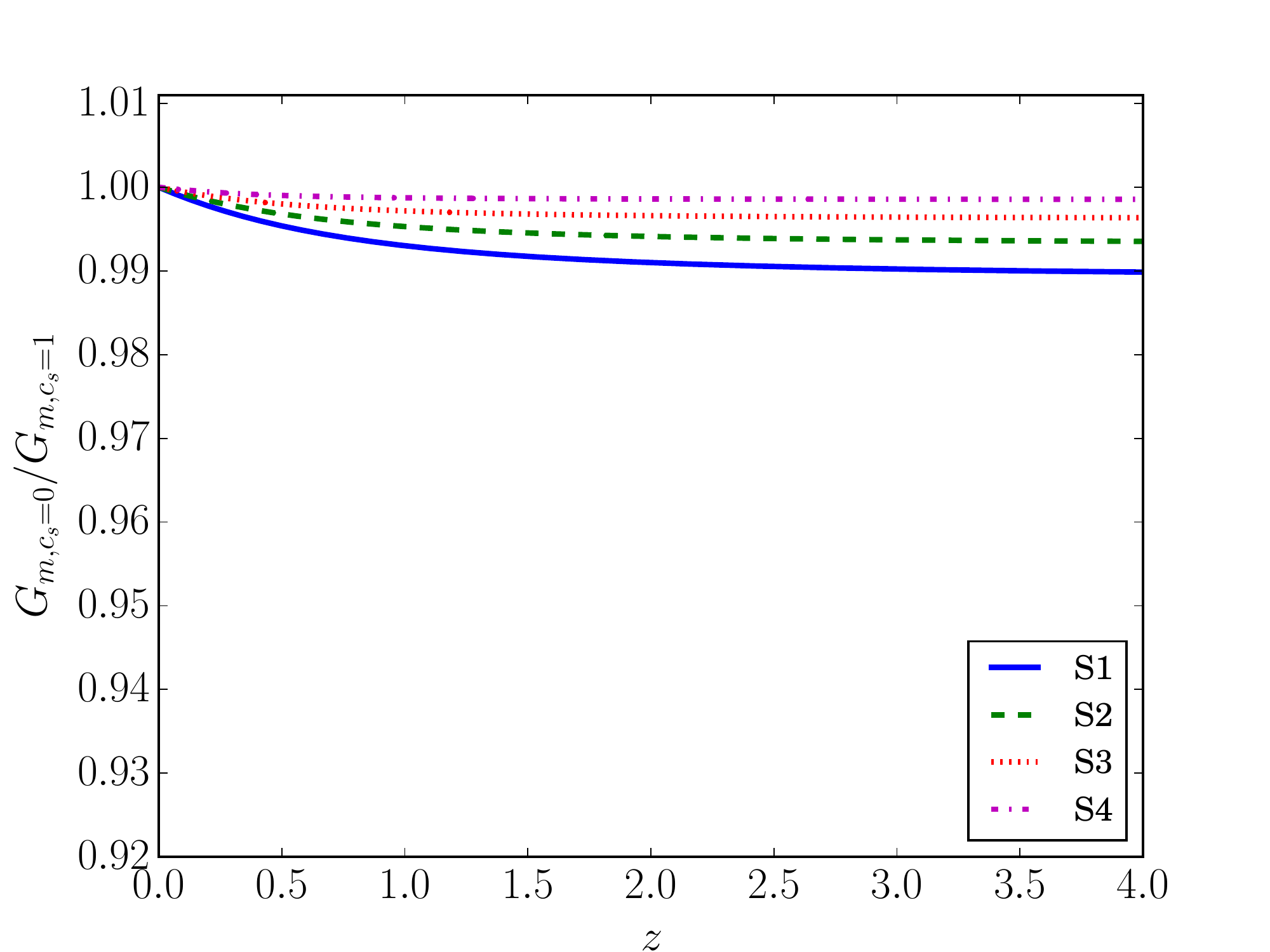}
\protect\caption{
Ratios $G_{{\rm tot},c_s=0}/G_{m, c_s=1}$ (left panel) and  
$G_{m,c_s=0}/G_{m, c_s=1}$ (right panel) as a function
of redshift for each set of parameters given in Table~\ref{tab:params}. 
\label{fig:growth}}
\end{figure}
\par\end{center}

\subsection{Virialization} \label{virialization}

Now we consider the nonlinear evolution of the fluctuations. First we
compute the virialization time, which will be used to define the critical 
density threshold for collapse, the virial mass and virial 
overdensity. 

If $c_{s}=0$, DE fluctuations have no pressure support and follow
matter trajectories. Their order of magnitude can be estimated using
Eq.~(\ref{eq:de_mat_relation}), which also shows that overdensities
in matter create overdensities in DE for $w>-1$. Moreover, as 
DE fluctuations grow in the nonlinear regime, its local equation,
$w_{c}=\left(\bar{\rho}+\delta\rho\right)/\left(\bar{p}+\delta 
p\right)=w\left[1-\delta_{de}/\left(1+\delta_{de}\right)\right]$,
tends to zero for $w>-1$, thus indicating that its properties become
more similar to matter, see~\cite{Mota:2004pa,Abramo2008}. 

Therefore, the total mass, $M_{{\rm tot}}$, is expected to have an extra 
contribution due to DE:
\begin{equation}
M_{{\rm tot}}=M_{m}+M_{{\rm de}}\,.
\end{equation}
The mass associated with matter is 
\begin{equation}
M_{m}=\frac{4\pi}{3}R^{3}\bar{\rho}_{m}\left(z\right)\left[1+\delta_{m}
\left(z\right)\right]\,,
\end{equation}
and is conserved. It is very debatable how to include DE contribution
from a phenomenological point of view. Since only DE fluctuations
with negligible sound speed are subjected to the same forces that act
on matter particles, we define 
\begin{equation}
M_{de}=\frac{4\pi}{3}R^{3}\bar{\rho}_{de}\left(z\right)\delta_{de}
\left(z\right)\,,
\end{equation}
which is not conserved throughout the evolution. Note that we are not 
including the contribution from the homogeneous energy density $\bar{\rho}_{de}$ 
(that we did include in the case of $M_m$).
This is choice is also helpful to connect to the case of 
homogeneous dark energy, whose uniform contribution is not usually taken into 
account when calculating the halo mass.

In order to determine the virialization time, we use the method described in 
Ref.~\cite{Basse2012},
which takes into account the non-conservation of $M_{{\rm tot}}$
caused by the inclusion $M_{de}$. In this approach the virialization
time is defined when the moment of inertia of a sphere of non-relativistic
particles is in steady state, which yields the equation
\begin{equation}
\frac{1}{2M_{{\rm tot}}}\frac{d^{2}M_{{\rm tot}}}{dt^{2}}+\frac{2}{M_{{\rm 
tot}}R}\frac{dM_{{\rm 
tot}}}{dt}\frac{dR}{dt}+\frac{1}{R^{2}}\left(\frac{dR}{dt}\right)^{2}+\frac{1}{R
}\frac{d^{2}R}{dt^{2}}=0\,.\label{eq:virial}
\end{equation}
In EdS, $R_{{\rm vir}}=\frac{1}{2}R_{{\rm ta}}$, and $\delta_{m}\left(z_{{\rm 
v}}\right)\simeq145.8$,
see Ref.~\cite{Lee2010b} for a discussion about this quantity and
its impact on the mass function.

\begin{figure}
\centering{}\includegraphics[width=1.0 
\columnwidth]{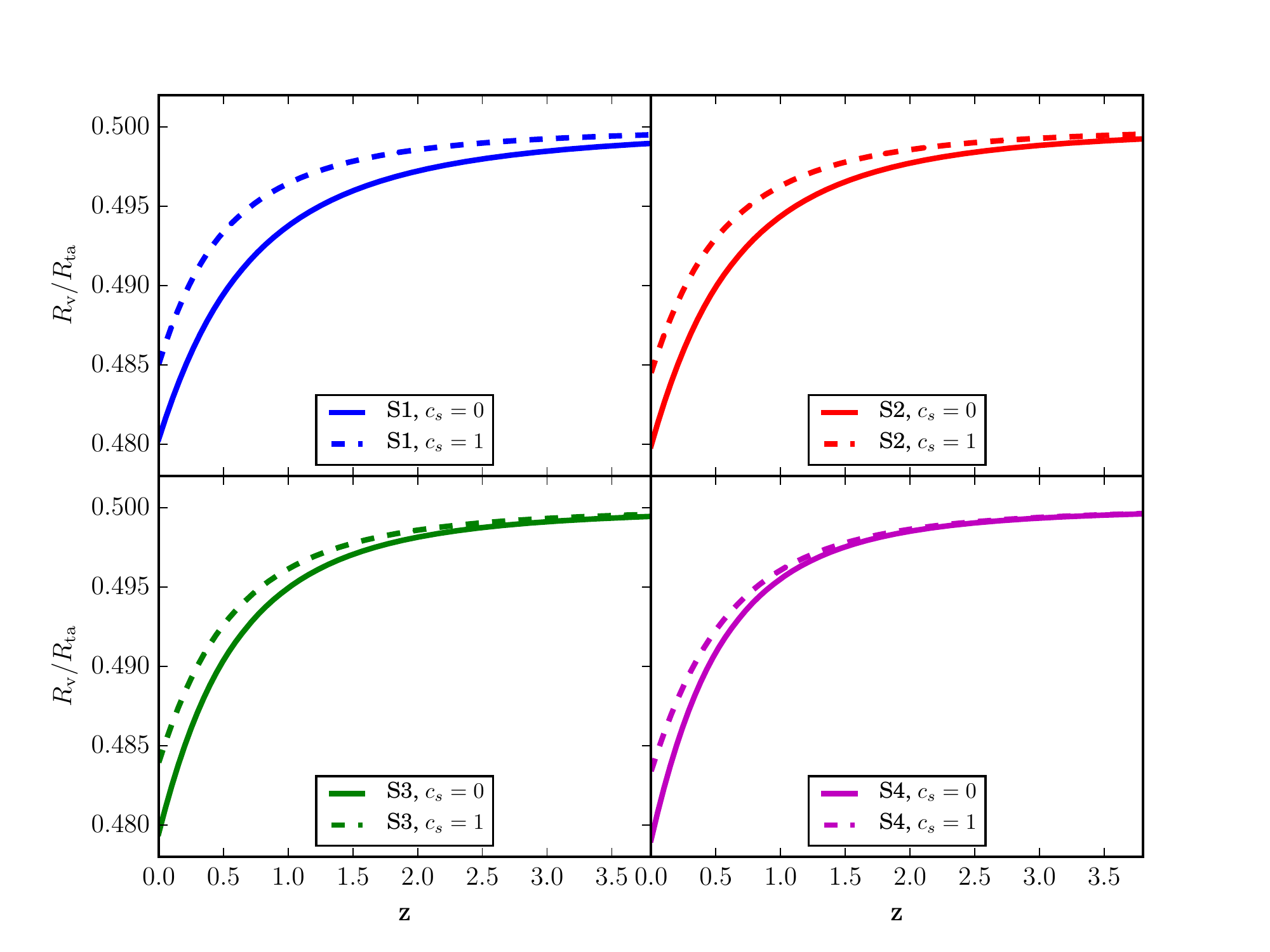}\protect\caption{$R_{{\rm v}}/R_{{\rm 
ta}}$ as a function of the virialization redshift for the set of parameters 
given in Table~\ref{tab:params}.
\label{fig:r_frac}}
\end{figure}

In Fig.~\ref{fig:r_frac} we show $R_{{\rm v}}/R_{{\rm ta}}$ for
each set of parameters and for $c_{s}=0$ and $c_{s}=1$.
DE fluctuations affect both the time variation of $M_{\rm tot}$ and $R$.
The general modification is that time derivatives
of $M_{\rm tot}$ delay the moment of virialization, so $R_{{\rm v}}/R_{{\rm 
ta}}$ is smaller when DE fluctuations are present. 

\begin{figure}
\centering{}
\includegraphics[width=.48 \columnwidth]{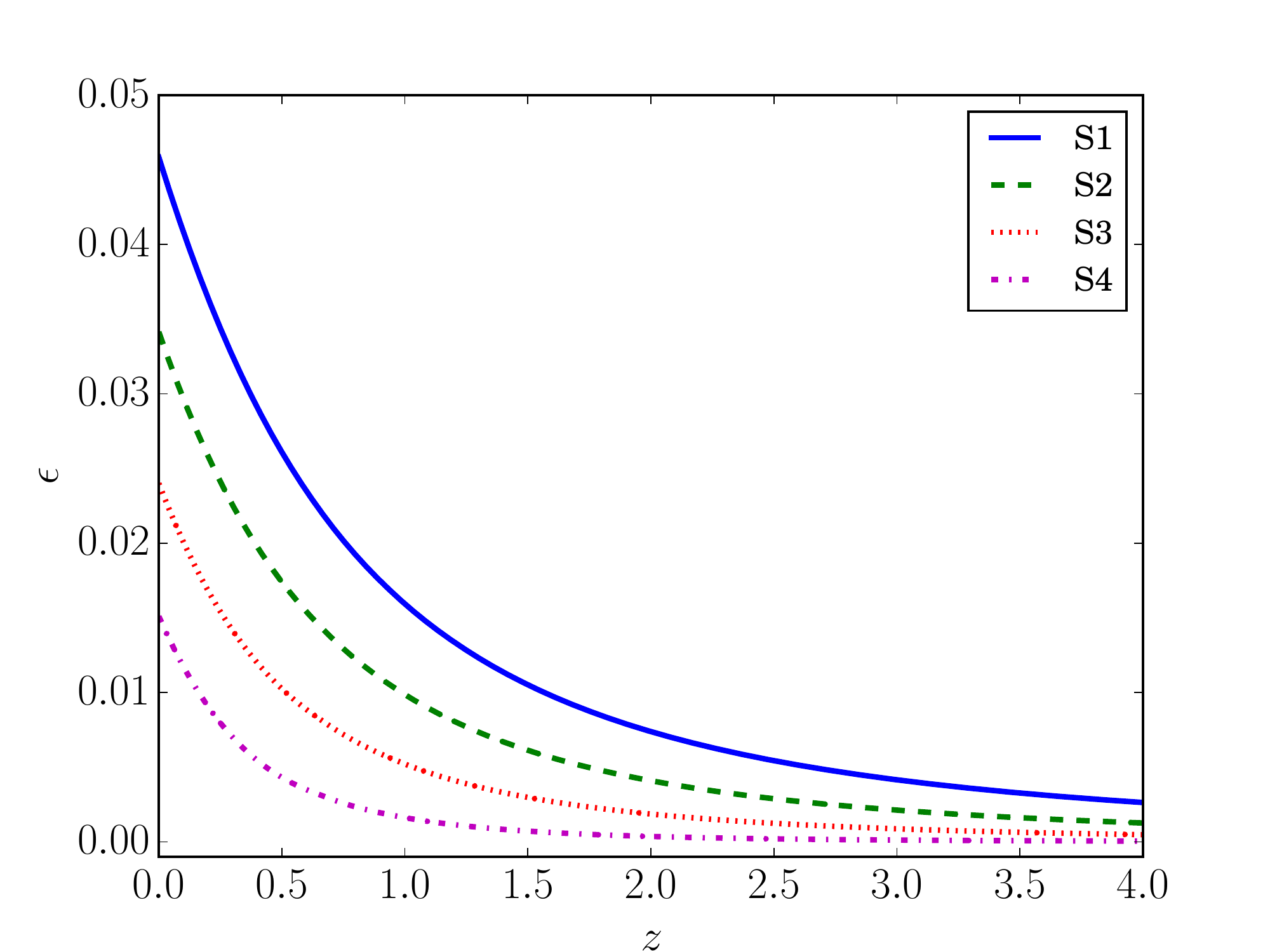}
\hfill
\includegraphics[width=.48 \columnwidth]{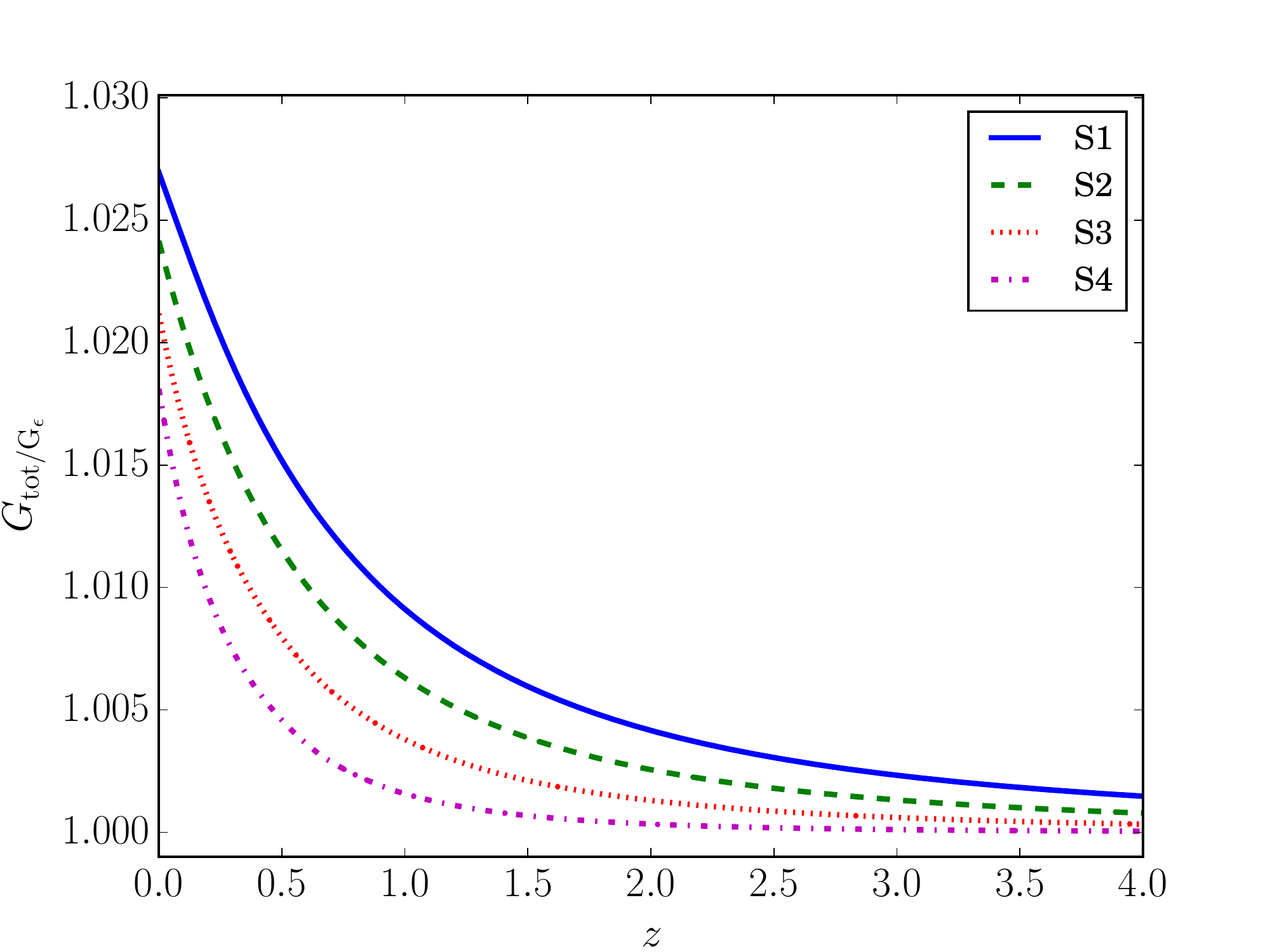}
\protect\caption{
{\it Left Panel:} Fraction of DE mass with respect to matter mass, 
$\epsilon=M_{de}/M_{m}$,
as a function of the virialization redshift for the set of parameters given in 
Table~\ref{tab:params}.
{\it Right Panel:} Plot of $G_{\rm tot}/ G_{\epsilon}$, see 
Section~\ref{virialization}.
\label{fig:epsilon}}
\end{figure}

In Fig.~\ref{fig:epsilon} (left panel) we show 
\begin{equation} \label{epsi}
\epsilon \equiv \frac{M_{de}\left(z_{{\rm 
v}}\right)}{M_{m}\left(z_{\rm{ v}}\right)}
\end{equation}
as a function of virialization redshift. This quantity is important because, as
we will show, it is related to the normalization of mass functions and also 
represents the fraction of DE mass in the halo.
It is clear that the contribution of DE increases as DE dominates
the background evolution. Since $M_{de}$ gives rise to the non-conservation
of total mass, the largest impact of DE fluctuations on quantities
computed at virialization is present at very low redshifts. It is
important to note that, the larger is $1+w$ at high-$z$,
the larger is $\epsilon$. As we can see, DE fluctuations can account for a few 
\% of the total mass of the halo.

As $\epsilon$ gives us the fractional contribution of DE with respect to matter, 
it should be linked to the contribution of the DE perturbation to $G_{\rm tot}$. 
It turns out that the function
\begin{equation} \label{G_epsilon}
G_{\epsilon}= G_{m,c_s=0} \left(1 + \epsilon\right)\, ,
\end{equation}
where $G_m$ is the matter growth function in the presence of clustering DE,
is very similar to $G_{\rm tot}$.  As one can see from Fig.~\ref{fig:epsilon} 
(right panel), these two quantities differ at most by $2.6\%$ at very low 
redshifts. This is a important consistency check between linear and nonlinear 
quantities that we have defined in this work.

\begin{figure}
\centering{}\includegraphics[width=1 \columnwidth]{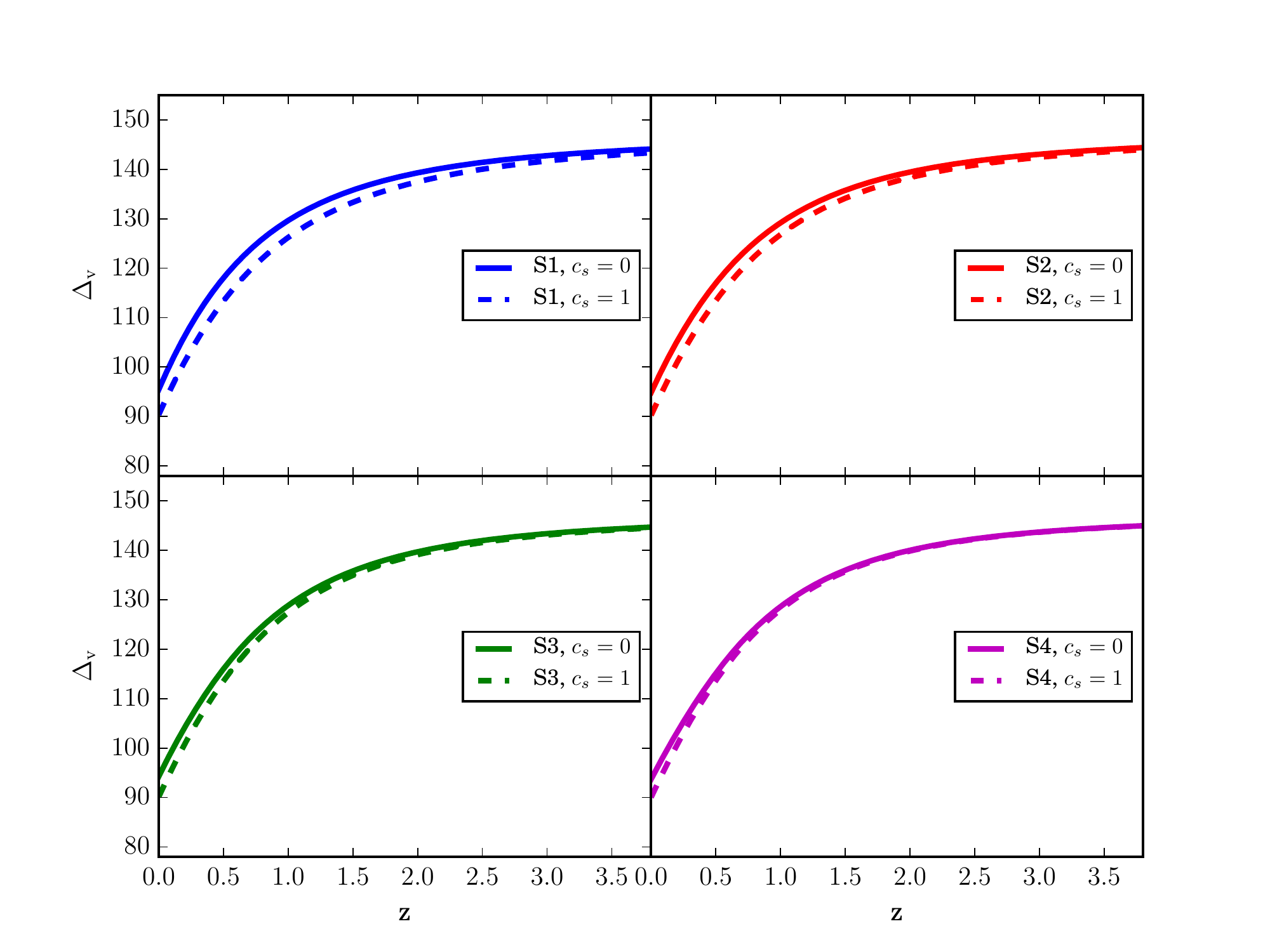}
\protect\caption{
Virialization overdensity $\Delta_{{\rm v}}$ of Eq.~(\ref{eq:DV_def}) as 
function of redshift
for the four sets of $w_{0}$ and $w_{a}$ given in Table \ref{tab:params}.
\label{fig:DV}}
\end{figure}

Expressing the total mass of the halo in the form $M_{{\rm 
tot}}=\frac{4\pi}{3}R_{{\rm vir}}^{3}\bar{\rho}_{c}\Delta_{{\rm v}}$,
the virialization overdensity is given by
\begin{equation}
\Delta_{{\rm v}}=\Omega_{m}\left(z_{{\rm 
v}}\right)\left[1+\delta_{m}\left(z_{{\rm 
v}}\right)\right]+\Omega_{de}\left(z_{{\rm v}}\right)\delta_{de}\left(z_{{\rm 
v}}\right)\,.\label{eq:DV_def}
\end{equation}
In Fig.~\ref{fig:DV} we show the evolution of $\Delta_{{\rm v}}$
as a function of redshift for the four different EoS evolution studied.
In all cases $\Delta_{{\rm v}}$ becomes smaller at low-$z$, as $\Omega_{m}$
decreases. The presence of DE fluctuations increases $\Delta_{{\rm v}}$,
which is a consequence of the delayed virialization that we discussed in the 
case of
$R_{{\rm v}}/R_{{\rm ta}}$.

\subsection{Critical density threshold} \label{critcon}

Usually PS and ST mass functions make use of $\delta_{c}$ as the critical 
density threshold that defines collapsed regions. However, when DE 
perturbations are important, mass is not conserved and the sensible choice 
is to define the halo properties at the virialization time.
Consequently, it seems inappropriate to use $\delta_{c}$
for the collapse threshold as in this case the mass function would depend on 
properties determined at the two different redshifts $z_{c}$ and $z_{{\rm v}}$. 
Therefore, we believe that the most natural quantity to use in the presence of 
DE fluctuations is the extrapolated linear contrast  $\delta_{{\rm tot}}^{{\rm 
lin}}$ at the moment of virialization,
\begin{equation}
\delta_{{\rm v}}\left(z_{{\rm v}}\right)=\delta_{{\rm tot}}^{{\rm 
lin}}\left(z_{{\rm v}}\right)\,.
\label{eq:delta_vir}
\end{equation}
which we define as $\delta_{{\rm v}}$ in order to avoid confusion with the
usual $\delta_{c}$. In Sec.~\ref{shmf} we show that this change can be absorbed
into a redefinition of one parameter of the ST mass function.

\begin{figure}
\centering{}\includegraphics[width=1 \columnwidth]{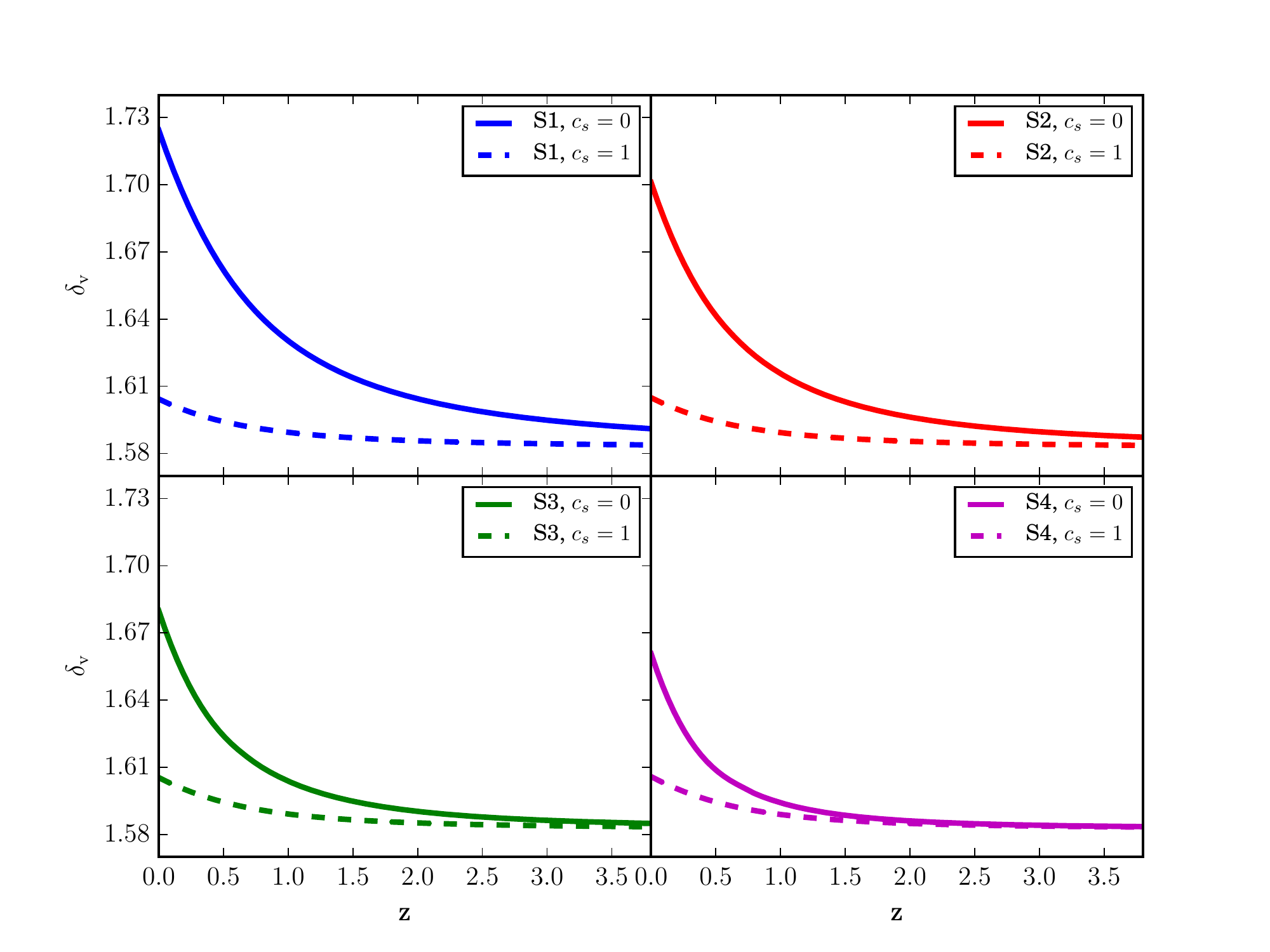}
\protect\caption{
Critical contrast at virialization, $\delta_{{\rm v}}$, given by 
Eq.~(\ref{eq:DV_def}), as a function of redshift for the four sets of $w_{0}$ 
and $w_{a}$ given in Table~\ref{tab:params}. See Section~\ref{critcon} for more 
details.
\label{fig:dc}}
\end{figure}

In Fig.~\ref{fig:dc} we show the evolution of $\delta_{{\rm v}}$
as a function of $z$. As for the other quantities, the largest impact
of DE fluctuations occur for S1. Note that, for both homogeneous and clustering 
DE, $\delta_{{\rm v}}$ grows as function of $z$, which is opposite to the usual
behavior of $\delta_{c}$ (see, 
e.g.,~\cite{Pace2010,Creminelli2010,Basse2011,Batista:2013oca}). The impact of 
DE fluctuations in $\delta_{{\rm v}}$ can be as large as 7\% for S1, whereas the 
impact in the usual $\delta_{c}$ is at most 1\%~\cite{Batista:2013oca}. These 
results are valid only for DE with  negligible $c_s$. For some range of small 
but non-negligible $c_s$, DE fluctuations are not negligible and can be treated 
linearly, \cite{Basse2011}. In this case, the impact of DE fluctuations is 
smaller, but induces mass dependence, $\delta_{c}(z)\rightarrow\delta_{c}(z,M)$, 
which is not considered in this work. 

\section{Halo mass function} \label{shmf}

The halo mass function $f(M,z)$ gives the fraction of the total mass
in halos of mass $M$ at redshift $z$. It is related to the (comoving)
number density $n(M,z)$ by: 
\begin{equation}
n(M,z)dM=\frac{\rho_{cc}(z)}{M}f(M,z)dM\,,
\label{hmf}
\end{equation}
where $\rho_{cc}$ is the comoving average of the energy density that makes up 
the collapsed halos.

In the usual case of homogeneous DE $\rho_{cc}(z)=\rho_{mc}\equiv 
a^{3}\bar{\rho}_{m}$,
where the latter is the constant matter density in a comoving volume
and $a$ is the background scale factor.

The halo mass function acquires an approximate universality when expressed
with respect to the variance of the mass fluctuations on a comoving
scale $r$ at a given redshift $z$, $\Delta(r,z)$. The latter is
given by: 
\begin{equation}
\Delta^{2}(r,z)\,=\,\int_{0}^{\infty}\frac{dk}{k}\tilde{\Delta}^{2}(k,z)W^{2}
(kr)\,,
\end{equation}
where $W(kr)$ is the Fourier transform of the top-hat window function
and $\tilde{\Delta}^{2}(k,z)$ is the dimensionless power spectrum
extrapolated using linear theory to the redshift $z$: 
\begin{equation}
\tilde{\Delta}^{2}(k,z)\equiv\frac{k^{3}}{2\pi^{2}}P(k,z)=\delta_{H0}^{2}
\left(\frac{ck}{H_{0}}\right)^{3+n_{s}}T^{2}(k)\, G_{{\rm 
tot}}^{2}(z)\,.\label{linearpk}
\end{equation}
In the previous equation $P(k,z)$ is the power spectrum, $n_{s}$
is the spectral index, $\delta_{H0}$ is the amplitude of perturbations
on the horizon scale today, $G_{{\rm tot}}(z)$ is the normalized
linear growth function given in Eq.~\eqref{eq:growth_N} and $T(k)$ is the
transfer function \cite{Eisenstein1998}. Note that the use of $G_{{\rm tot}}(z)$ 
accounts for the contribution of DE to the power spectrum. Since we 
are considering only the cases of $c_{s}=1$ and $c_{s}=0$, the transfer 
function shape should remain the same because DE perturbations are either 
negligible or have the same scale dependence of matter perturbations, thus 
changing only the overall normalization which is fixed via 
\begin{equation}
\Delta(r=8h^{-1}\text{Mpc},z=0)=\sigma_{8}\,,
\end{equation}
where $\sigma_{8}$ takes the fiducial value given earlier. Note that we are 
normalizing all the models to the same present-day value of $\sigma_8$.
For the impact of general $c_{s}$ in the power spectrum see~\cite{Kunz2015}. 

We model the mass function according to the functional form proposed
by Sheth and Tormen (ST)~\cite{Sheth1999}: 
\begin{equation}
f_{{\rm 
ST}}(\Delta)=A\sqrt{\frac{2}{\pi}}\left[1+\left(\frac{\Delta^{2}}{a\delta_{c}^{2
}}\right)^{p}\right]\frac{\sqrt{a}\delta_{c}}{\Delta}\exp\left[-\frac{a\delta_{c
}^{2}}{2\Delta^{2}}\right].\label{st}
\end{equation}
In the previous equation $a$ and $p$ are free parameters while $A$
is fixed by the normalization condition $\int fdM=1$, according to which: 
\begin{equation}
A=\left[1+\frac{2^{-p}}{\sqrt{\pi}}\Gamma(1/2-p)\right]^{-1}\,.\label{Aofp}
\end{equation}
We adopt here the original values from \cite{Sheth1999}, which are
$a=0.707$ and $p=0.3$ and are expected to be approximately universal (see, 
however, \cite{Castro:2016jmw}).
In the case of clustering DE this is justified by the fact that for 
$c_s=0$ the shape of the power spectrum is not changed.
However, we should point out that, to our knowledge, a numerical simulation 
that includes clustering DE has not yet been run and, therefore, the 
ST parameters $a$ and $p$ may depend on the DE properties. A possible 
change will likely increase the impact of inhomogeneous DE on the halo 
abundances. Consequently, by neglecting this effect our results will be more 
conservative.

We adopted the ST mass function as it depends on the peak height parameter $\nu = 
\delta_c/\Delta$ and not on $\Delta$ alone, as in the case of the precise mass 
function proposed by \cite{Tinker2008b}. This makes the ST function more 
sensitive to the physics that goes into the halo collapse that we are studying 
in this paper.
In order to obtain a more precise mass function, the usual approach
-- adopted, for example, in the context of non-Gaussianities \cite{LoVerde2008}
and baryonic feedback \cite{Velliscig2014} -- is to multiply the
more precise but less cosmology-dependent mass functions by a correcting
factor, which in our case is: 
\begin{equation}
\frac{n(M,z)|_{c_{s}=0}}{n(M,z)|_{c_{s}=1}}=\frac{\rho_{cc}(z)f(M,z)|_{c_{s}=0}}
{\rho_{mc}\, f(M,z)|_{c_{s}=1}}\,.
\end{equation}
We shall then focus on this correcting factor when presenting our results.

As previously discussed, we define halos' properties at the moment
of virialization. The ST mass function, however, uses the critical contrast at 
collapse, $\delta_{c}$.
Therefore, we need to adjust the parameter $a$ -- which is degenerated with 
$\delta_{c}$ -- in order to account for the different critical contrast used:
\begin{equation}
\tilde{a}\equiv a\frac{\delta_{c}^{2}}{\delta_{{\rm v}}^{2}}\simeq0.803 \,,
\end{equation}
where within the EdS mode it is $\delta_{\rm v} \simeq1.583$ and 
$\delta_c\simeq1.686$.
The ST mass function becomes: 
\begin{equation}
f_{{\rm 
ST}}(\Delta)=A\sqrt{\frac{2}{\pi}}\left[1+\left(\frac{\Delta^{2}}{\tilde{a}
\delta_{{\rm v}}^{2}}\right)^{p}\right]\frac{\sqrt{\tilde{a}}\delta_{{\rm 
v}}}{\Delta} 
\exp\left[-\frac{\tilde{a}\delta_{{\rm v}}^{2}}{2\Delta^{2}}\right].\label{st2}
\end{equation}

Note that, because of the change of variable, $f_{{\rm ST}}$ is related
to our original definition of $f$ by: 
\begin{equation} \label{jacob}
f(M,z)\,=\, f_{{\rm ST}}(\Delta)\frac{d\ln\Delta(M,z)^{-1}}{dM}\,.
\end{equation}
In the latter, we have $\Delta(M,z)=\Delta(r(M,z),z)$, where the function
$r(M,z)$ relates the comoving scale that collapsed to form the halo
with the actual halo mass. Within homogeneous DE this relation is
simply: 
\begin{equation}
M=\frac{4\pi}{3}r^{3}\rho_{mc}\left(1+\delta_{m}\right)=\frac{4\pi}{3}r^{3}a^{3}
\bar{\rho}_{m}\left(1+\delta_{m}\right)\,,\label{eq:mass_scale}
\end{equation}
that is, thanks to mass conservation, one can take the mass corresponding
to the collapsing sphere at early times when the perturbation is negligible
and $\rho_{m}\approx\bar{\rho}_{m}$. Therefore, one get a ``background level'' 
relation $r=r(M)$ and it is not necessary
to know the virialization time and the corresponding overdensity and
halo radius.

\begin{figure}
\centering{}\includegraphics[width=.6 
\columnwidth]{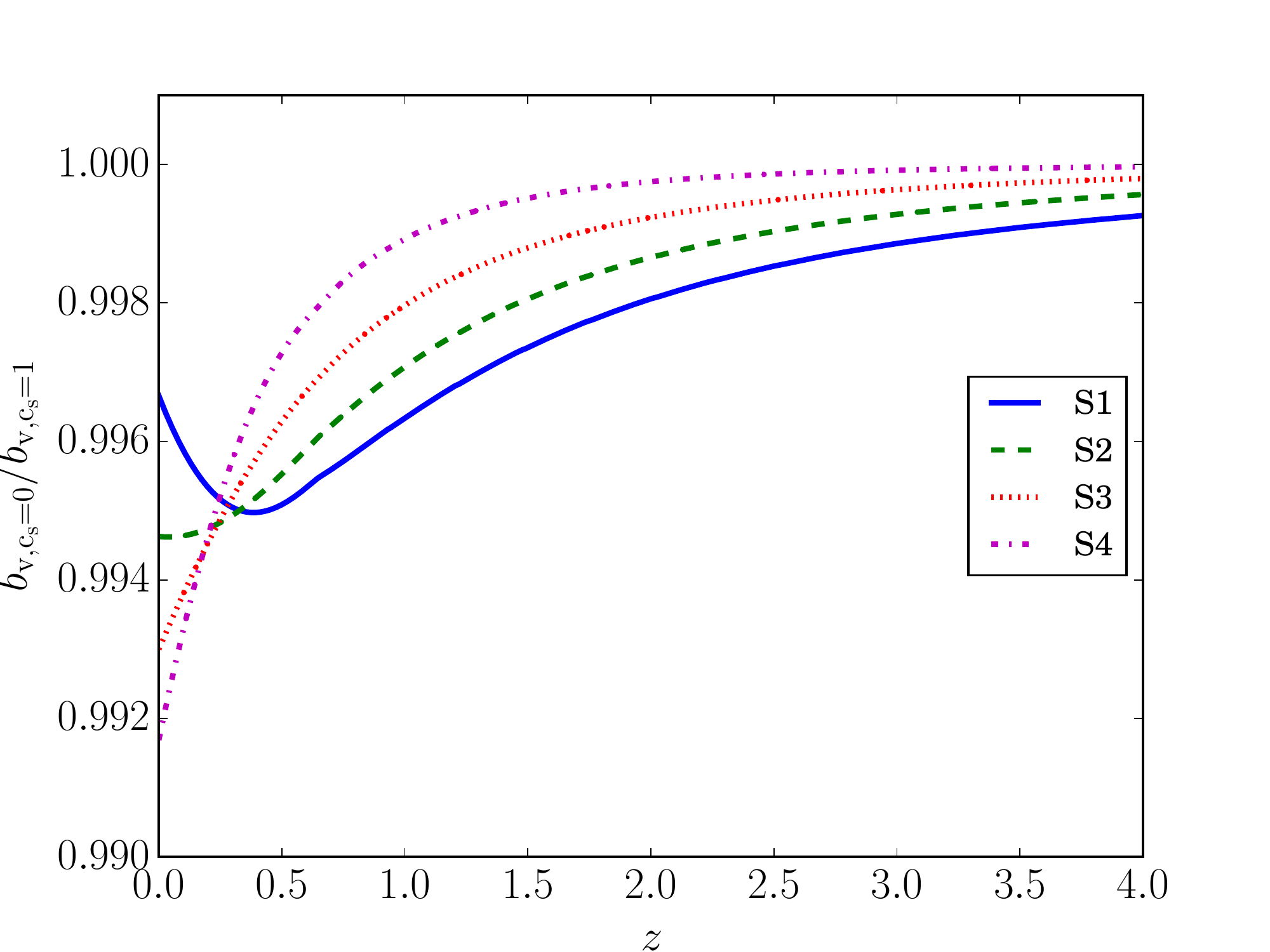}\protect\caption{Plot of the ratio 
$b_{{\rm v},c_{s}=0}/b_{{\rm v},c_{s}=1}$ as function
of redshift for the four sets of $w_{0}$ and $w_{a}$ given in 
Table~\ref{tab:params}. See Section~\ref{shmf} for more details.
\label{fig:bv}}
\end{figure}

As discussed in Section~\ref{virialization}, when dark energy perturbations are 
present the total mass $M_{{\rm tot}}$
is not conserved. Hence, we need to define the actual halo mass and the
radius at the virialization and the relation
$r(M,z)$ is given by: 
\begin{equation}
M_{{\rm tot}}=\frac{4\pi}{3}r^{3}b_{{\rm v}}^{3}(z)\Delta_{{\rm 
v}}(z)\bar{\rho}_{c}(z)\,,
\label{mass}
\end{equation}
where the physical halo radius is $R_{{\rm v}}\equiv r \, b_{{\rm v}}$,
$b_{{\rm v}}(z)$ is independent of the comoving scale $r$ and so
of the halo mass $M$. Considering $R_{i}=b_{i}r\simeq a_{i}r$, $b_{{\rm v}}$
can be computed using: 
\begin{equation}
b_{{\rm v}}=\frac{R_{{\rm v}}a_{i}}{R_{i}}\,.
\label{eq:bv_general}
\end{equation}
If only matter contributes to the mass, $M_{{\rm 
tot}}=M_{m}\left(z_{i}\right)=M_{m}\left(z_{{\rm v}}\right)$
is conserved and we can define the conserved $b_{{\rm v}}$ according to
\[
\frac{4\pi}{3}r^{3}\rho_{mc}\left(1+\delta_{m}\left(z_{i}\right)\right)=\frac{
4\pi}{3}r^{3}b_{{\rm v}}^{3}a^{-3}\rho_{mc}\left(1+\delta_{m}\left(z_{{\rm 
v}}\right)\right)
\]
so that
\begin{equation}
b_{{\rm v}}^{{\rm cons}}=a_{{\rm 
v}}\left(\frac{1+\delta_{m}\left(z_{i}\right)}{1+\delta_{m}\left(z_{{\rm 
v}}\right)}\right)^{1/3}\,,
\end{equation}
which is equal to the general case of Eq.~(\ref{eq:bv_general}) if
dark energy fluctuations are negligible. We note that the common practice of
neglecting $\delta_{m}\left(z_{i}\right)$ in (\ref{eq:mass_scale})
induces an error $\sim0.1\%$ in $b_{{\rm v}}^{{\rm cons}}$. In 
Fig.~\ref{fig:bv} we show the evolution of the ratio $b_{{\rm 
v},c_{s}=0}/b_{{\rm v},c_{s}=1}$
for each set of parameterizations. In all cases, the modifications with
respect to $b_{{\rm v}}^{{\rm cons}}$ grow with redshift.
The largest differences occur for S3 and S4, a direct consequence of the impact 
of DE fluctuations on $R_{{\rm v}}$ as shown in Fig.~\ref{fig:r_frac}.

\begin{figure}
\begin{center}
\includegraphics[width=.47 \columnwidth]{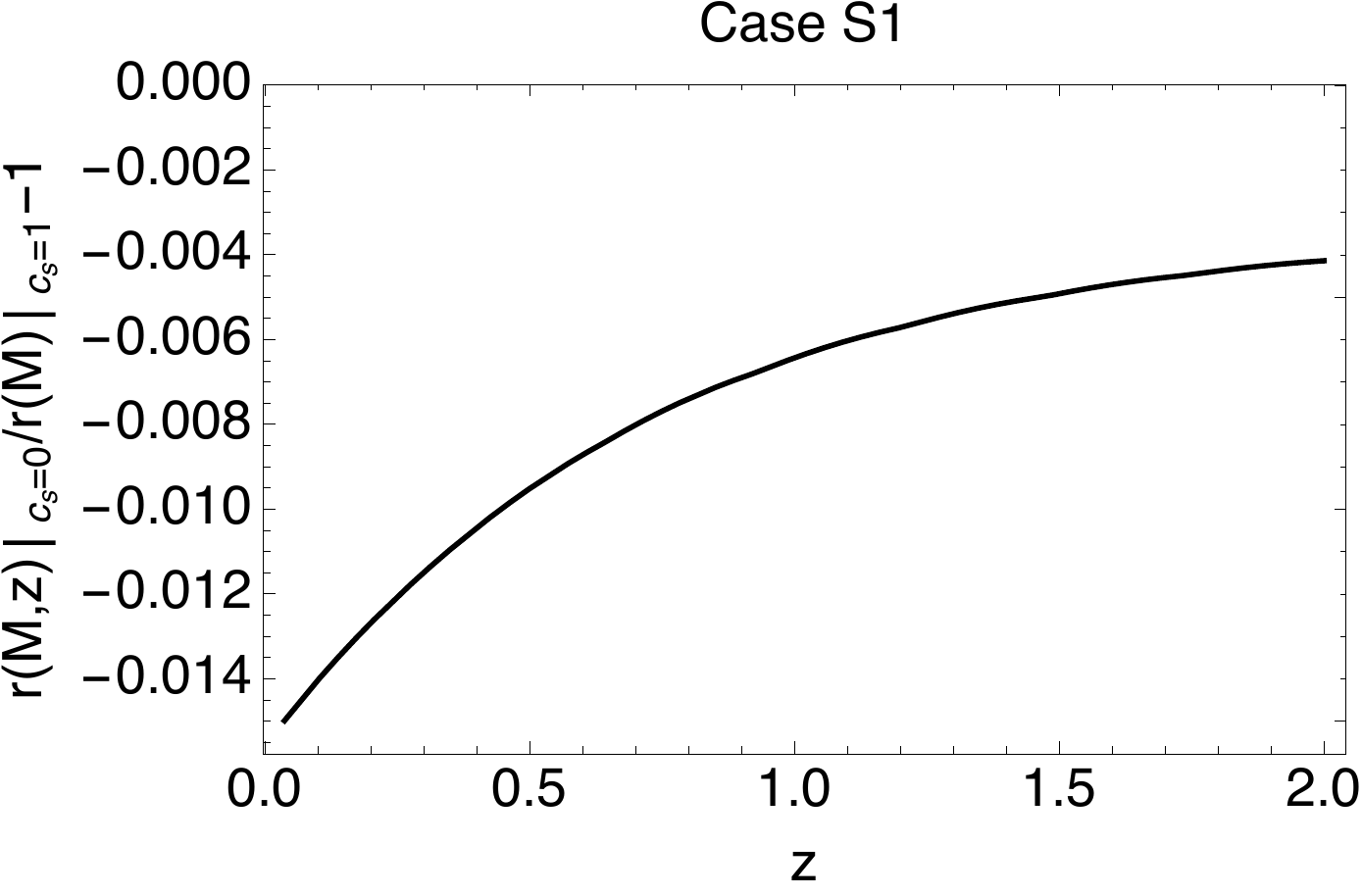}
\hfill
\includegraphics[width=.48 \columnwidth]{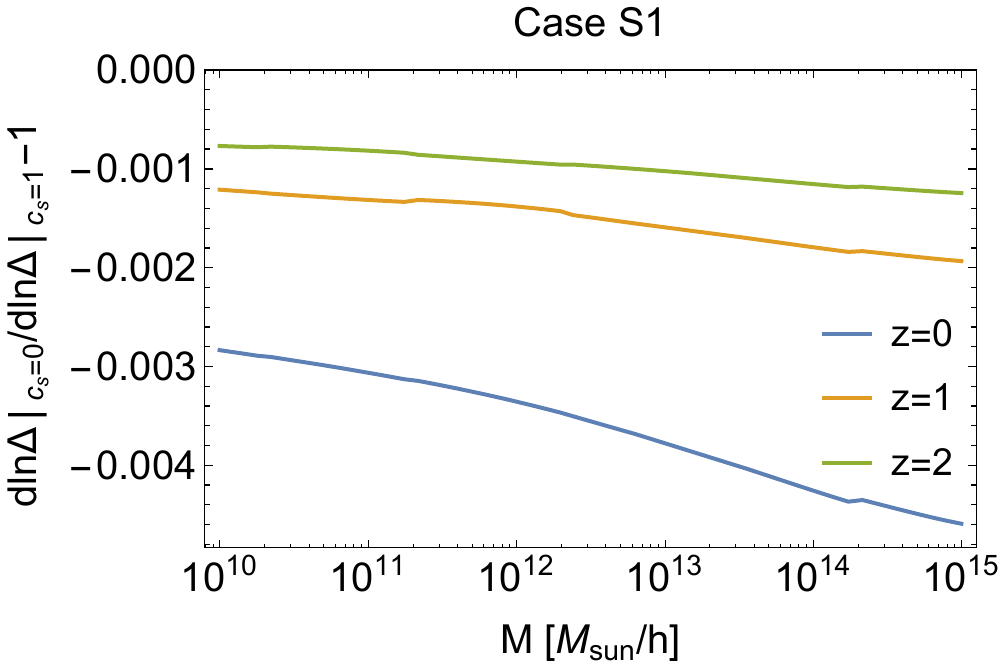}
\caption{
{\it Left Panel:} Fractional difference of the non-standard $r=r(M,z)$ relation 
from equation \eqref{mass} with respect to the background relation $r=r(M)$.
{\it Right Panel:} 
Impact of the non-standard $r=r(M,z)$ relation on the logarithmic derivative of 
\eqref{jacob}.
These plots refer to the set S1 given in Table~\ref{tab:params}. 
}
\label{dlnD-Case_S1-norm_1}
\end{center}
\end{figure}

Fig.~\ref{dlnD-Case_S1-norm_1} shows how different is the non-standard 
$r=r(M,z)$ relation with respect to the background relation $r=r(M)$ (left 
panel) and the impact of the non-standard $r=r(M,z)$ relation on the 
logarithmic derivative of \eqref{jacob}. Here we only show the results for the case S1, 
which suffers the largest modifications. As one can see, the impact of clustering DE 
in the mass-scale relation is around 1\% at $z=0$ (left panel of Fig.~\ref{dlnD-Case_S1-norm_1}), however the actual impact on the mass function is 
one order of magnitude smaller because it depends on a logarithmic derivative 
(right panel of Fig.~\ref{dlnD-Case_S1-norm_1}) and can be safely neglected.

%
%%%%%%%%%%%%%%%%%%%%
%%%%%%%%%%%%%%%%%%%%
%

Finally, the last quantity that we need to evaluate is $\rho_{cc}$.
The total average mass density in halos at redshift $z$ can be obtained from 
the following equation:
\begin{align}
\rho_{cc}(z) &= \int M \, n(M,z) \rd M =  \int M_m \, n(M,z) \rd M +  \int 
M_{de} \, n(M,z) \rd M \nonumber \\
&=  \int M_m \, n(M,z) \rd M +  \epsilon(z) \int M_{m} \, n(M,z) \rd M = 
\rho_{mc} \, [1+\epsilon(z)] \,,
\label{eq:rho_normalization}
\end{align}
where it has been assumed that, at redshift $z$, all halos have the same ratio 
$\epsilon(z)$ given in Eq.~\eqref{epsi}.
This neglects the fact that large halos are produced by the mergers of smaller 
halos that have virialized earlier and so with a different virialization 
overdensity and DE contribution. As we can see, the impact on the mass function 
is linear in $\epsilon$ (left panel of Fig.~\ref{fig:epsilon}), so it increases 
abundances as much as 5\% on all mass scales.

\section{Results} \label{results}

%
%%%%%%%%%%%%%%%%%%%%
\begin{figure}
\begin{center}
\includegraphics[width=.48 \columnwidth]{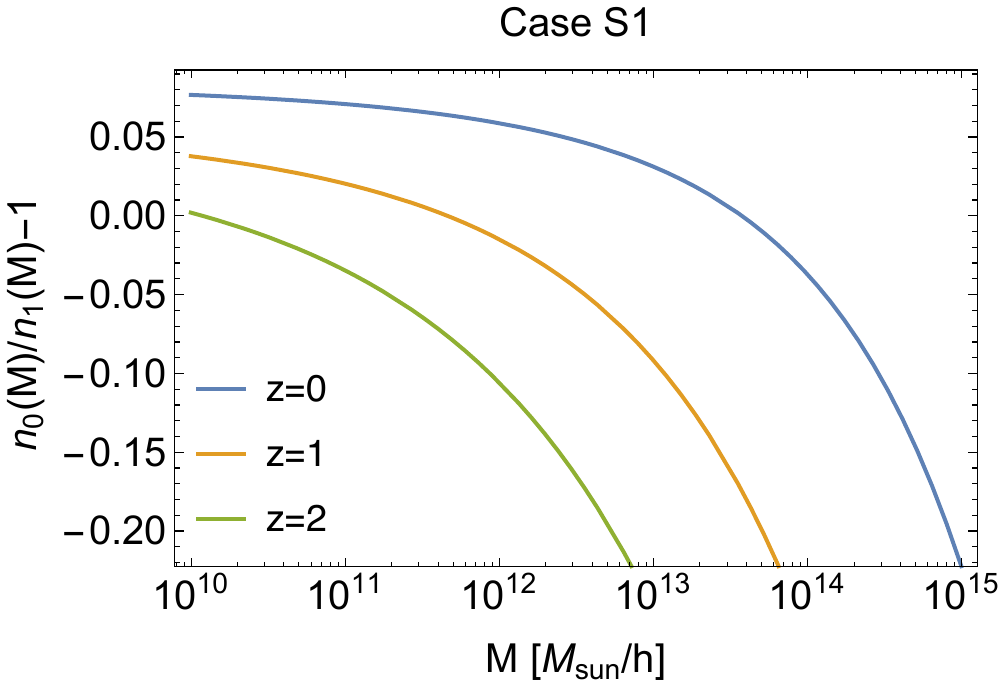}
\hfill
\includegraphics[width=.48 \columnwidth]{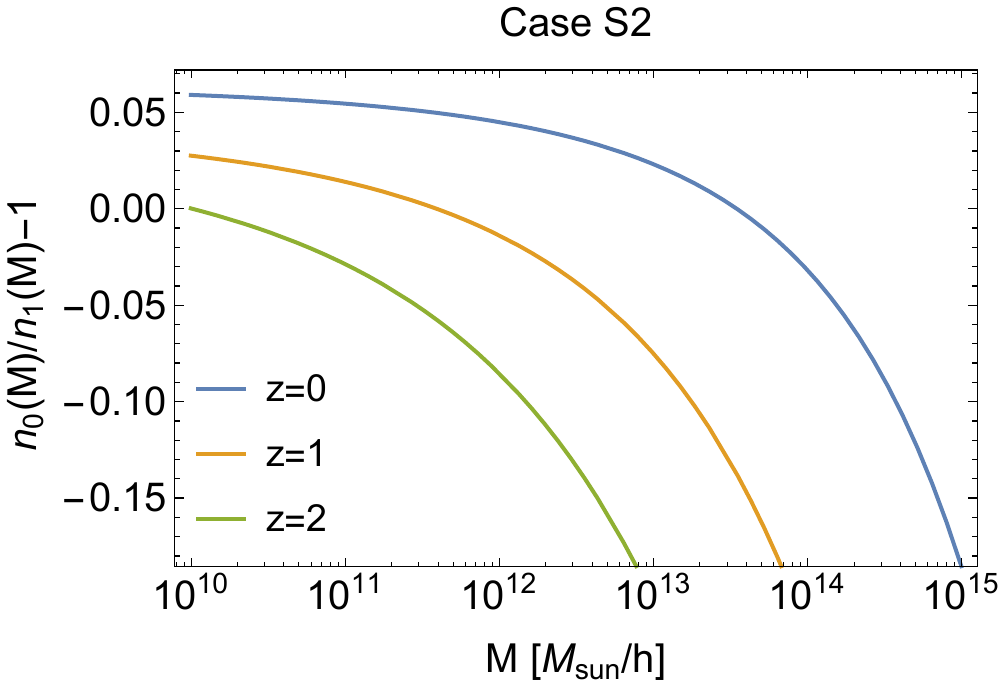}\\
\vspace{.7 cm}
\includegraphics[width=.48 \columnwidth]{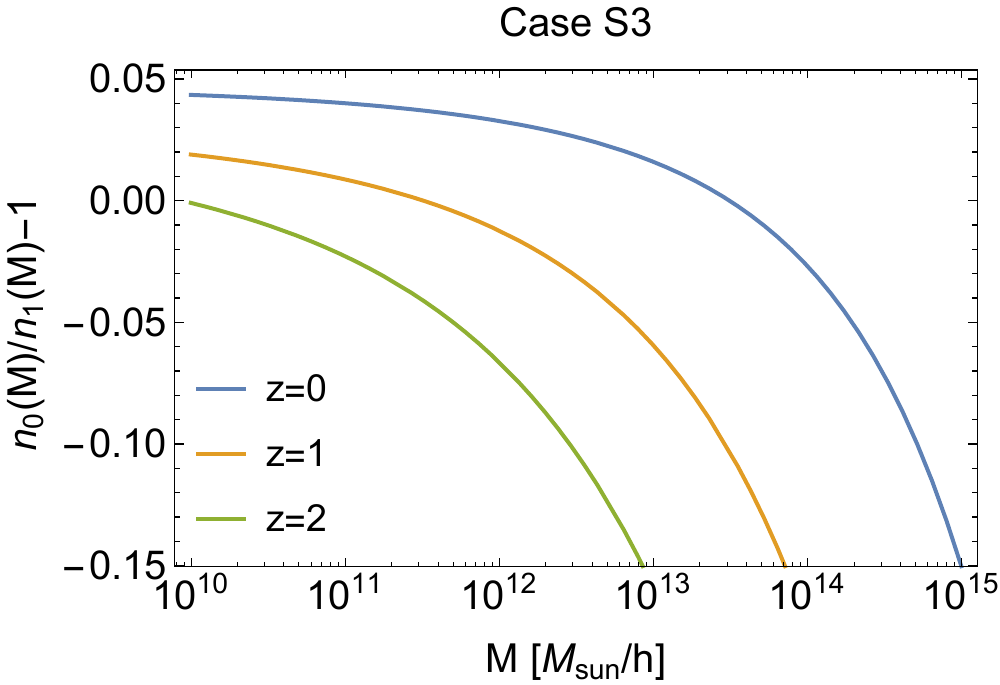}
\hfill
\includegraphics[width=.48 \columnwidth]{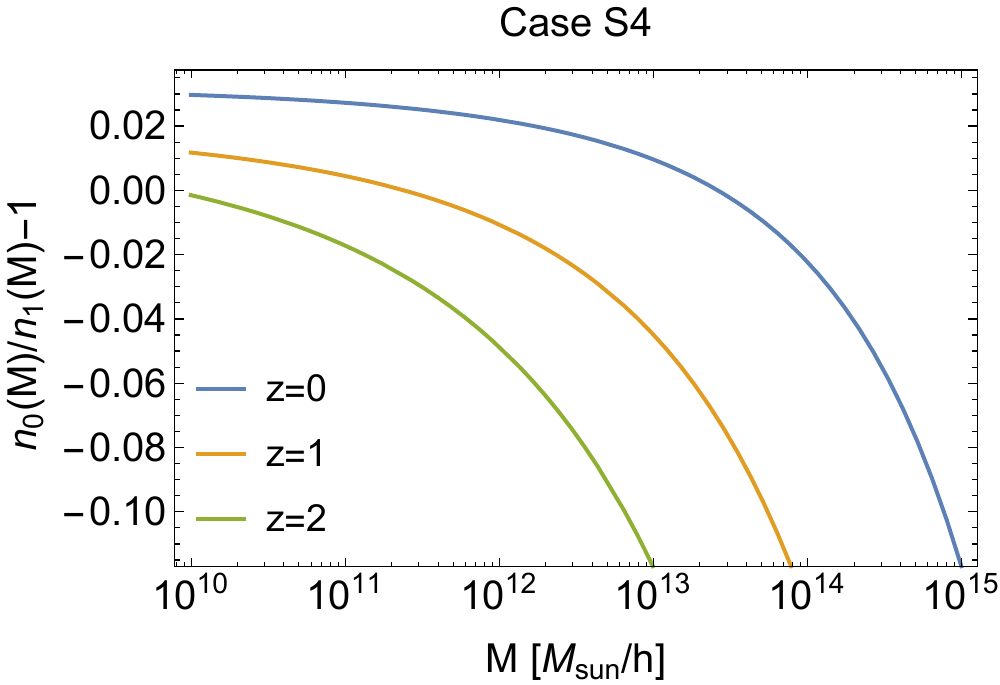}
\caption{
Fractional change in the mass function due to dark energy perturbations.
$n_0(M)$ refers to dark energy with sound speed $c_s=0$, and similarly for 
$n_1(M)$.
Perturbations are normalized today so that all models have the same $\sigma_8$.
The four sets S1-4 of $w_{0}$ and $w_{a}$ are given in Table~\ref{tab:params}. 
}
\label{nratio-norm1}
\end{center}
\end{figure}
%%%%%%%%%%%%%%%%%%%%
%

We show in Fig.~\ref{nratio-norm1} the quantity 
$\frac{n(M,z)|_{c_{s}=0}}{n(M,z)|_{c_{s}=1}}$ for the four sets of $w_{0}$ 
and $w_{a}$ given in Table~\ref{tab:params} and at three different redshifts. 
As one can see, DE perturbations can impact halo abundances at the 
level of 10\%--30\%, depending on the halo mass and redshift. Bear in mind that 
all models have the same $\sigma_8$. In this case, clustering DE 
always diminishes the abundance of massive clusters $M>10^{14} M_{\odot}$. 
Interestingly, the number of small halos,  $10^{10} M_{\odot}<M<10^{12} 
M_{\odot}$, is enhanced at $z=0$ and $z=1$. The main impact of clustering DE 
for large masses comes from the modifications in the quantity 
$\delta_{\rm{v}}/G_{{\rm tot}}$, which is always larger than in the homogeneous 
case and strongly reduces the number of halos around the exponential tail of the 
mass function. For small masses, the main impact is related to the density 
normalization of mass function, Eq.~\eqref{eq:rho_normalization}. 

Had we normalized the amplitude of fluctuations at the redshift 
of CMB decoupling, clustering DE would have generated more massive clusters in 
comparison to homogeneous DE, therefore worsening the Planck Clusters tension. 
Clustering DE models alleviate this tension only if $w<-1$, in which case the 
impact of DE fluctuations is opposite to the case $w>-1$. However, as we 
discussed in Sect. \ref{bck_mods}, phantom clustering DE can generate negative 
energy densities, $\delta_{de} < -1$. We will address this problem in a
forthcoming paper ~\cite{RoVaX}.   

There are numerous ongoing and planned surveys (e.g., 
DES~\cite{Abbott:2005bi,Abbott:2017wau}, 
J-PAS~\cite{Benitez:2014ibt,Ascaso:2016ddl}, 
Euclid~\cite{Laureijs:2011gra,Sartoris:2015aga,Amendola:2016saw}) which will 
count the number of high-mass systems on the sky in an effort to better 
constrain the cosmological parameters that control the growth rate of structure.
As we have seen the impact of DE perturbations on the mass function is 
greatest at large masses. Therefore, it is interesting to quantify how much 
cluster counts are affected by such perturbations.

Here, we neglect the effect of the mass-observable relation and we do not bin 
in mass. We will carry out survey-specific analyses in a forthcoming work.
We define a cluster as an object with mass $M \ge M_{\rm thr} \equiv 10^{14} 
h^{-1} M_{\odot}$, a figure in agreement with the expected sensitivity from 
Euclid. The total number of clusters at the redshift $z$ and within the 
redshift bin $\Delta z$ is then:
\begin{align} \label{Nz}
N(z) \Delta z=  \Delta z \frac{\rd V}{\rd z} \int_{M_{\rm thr}}^{\infty} \rd M 
\, n(M, z)  \,,
\end{align}
where the quantity $\rd V/\rd z$ is the cosmology-dependent comoving volume 
element per unit redshift interval which is given by:
\begin{equation}
\frac{\rd V}{\rd z} = 4 \pi  (1+z)^2 \, \frac{d_A^2(z)}{c^{-1} H(z)} \,,
\end{equation}
where $d_A$ is the angular diameter distance and $H(z)$ is the Hubble rate at 
redshift $z$.

\begin{figure}
\begin{center}
\includegraphics[width=.6 \columnwidth]{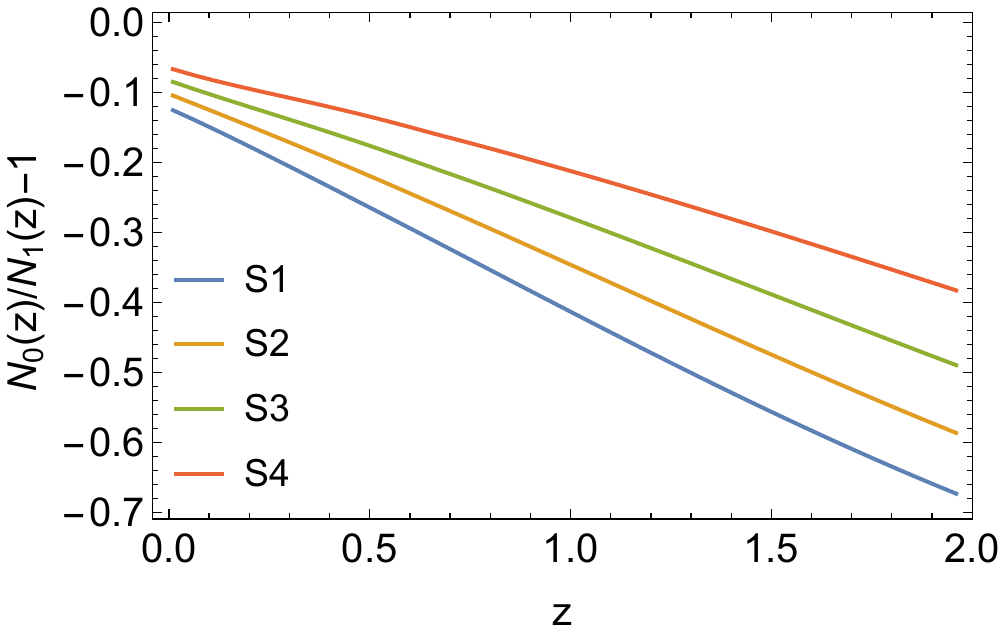}
\caption{Fractional change in cluster number counts due to dark energy 
perturbations.
$N(z)$ is computed using \eqref{Nz} and $N_0(z)$ refers to dark energy with 
sound speed $c_s=0$, and similarly for $N_1(z)$. Perturbations are normalized 
today so that all models have the same $\sigma_8$.
The four sets S1-4 of $w_{0}$ and $w_{a}$ are given in Table~\ref{tab:params}.}
\label{Nclratio-norm1}
\end{center}
\end{figure}

Figure~\ref{Nclratio-norm1} shows that cluster counts are modified by about 
30\% at a redshift of unity. Therefore, cluster counts is a very promising 
observable if one is to constrain models of dark energy that feature negligible 
sound speeds.

%
%%%%%%%%%%%%%%%%%%%%
%%%%%%%%%%%%%%%%%%%%
%

\section{Conclusions} \label{conclusions}

In this paper we have discussed in detail how the halo mass function is 
modified when dark energy has a negligible sound speed that causes its 
perturbations to collapse.
We have developed a consistent framework according to which the halo properties 
are defined at the moment of virialization and that takes into account the 
non-conservation of the mass relative to the dark-energy fluctuations. 

To summarize, the corrections that directly impact the mass function due 
to the presence of DE fluctuations are :
\begin{itemize}
\item Growth function: based on the spherical collapse model in the radius 
approach, we have argued that the total perturbation density should be used, 
Eqs.~\eqref{eq:delta_tot} and \eqref{eq:growth_N}. This new function differs 
from matter growth function in the homogeneous case by 3\% to 8\% for the equations of state we have analyzed. 
\item Threshold density for collapse: by demanding consistency in the 
definition of halo mass and the mass function quantities, we have argued that 
most natural threshold density to be used is the total linear perturbation 
density at the moment of virialization, $\delta_{{\rm v}}$, Eq.~\eqref{eq:delta_vir}. The impact of clustering DE on this quantity is similar 
to the one on the growth function, but has a distinct redshift evolution in 
comparison with the usual $\delta_c$.    
\item Mass-scale relation: since the total mass of a halo with some fraction of 
DE fluctuation is not conserved, one can not use the background mass-scale 
relation for matter only. The corrected relation is given by Eqs. \eqref{mass} 
and \eqref{eq:bv_general}. This, however, is a sub-percent correction and can be 
safely neglected. 
\item Normalization density: also because DE fluctuations induce 
non-conservation of total mass, the mean energy density that normalizes the 
mass function has to be corrected, Eq.~\eqref{eq:rho_normalization}. This is a 
linear modification of the order of the DE mass fraction $\epsilon$ and so it changes the 
mass function by a few \% on all mass scales. 
\end{itemize}

We have then computed the impact of clustering dark energy considering all the 
mentioned modifications on the mass function and found that halo abundances 
can be altered at the level of 10\%--30\%, depending on the 
halo mass and redshift. As the change is largest at the high masses of galaxy 
clusters, we have then computed the impact on cluster number counts and 
obtained that they are modified by about 30\% at a redshift of unity, as big an 
effect as the one that comes from including baryons within cosmological 
simulations~\cite{Velliscig2014}. A comprehensive analysis that takes into 
account the nuisances from the scaling relations will be the subject of 
forthcoming work~\cite{RoVaX}.

\acknowledgments

The authors would like to warmly thank Rogério Rosenfeld for his valuable 
contribution during the early stages of this work, and also the 
ICTP-SAIFR/IFT-UNESP for supporting the authors and promoting the beginning of 
this project.
The authors benefitted from discussions with Iggy Sawicki, Caroline Heneka, 
Jorge Noreña and Francesco Pace.
VM is supported by the Brazilian research agency CNPq. RBC thanks the 
International Institute of Physics of Rio Grande do Norte Federal University 
for the support in providing an adequate office. RCB is ever grateful to his 
doughter Alice for the two days that she enlightened his life.

\bibliographystyle{JHEP}
\bibliography{referencias}
%\bibliography{biblio,halo,cosmo-lensing,growth-rate}

\end{document}